\documentclass[aps,showpacs,preprintnumbers,amsmath,amssymb,nofootinbib,preprint,showkeys]{revtex4}

\usepackage{graphicx}
\usepackage{color}
      
\def \be  {\begin{equation}}
\def \ee  {\end{equation}}
\def \ee  {\end{equation}}
\def \bea {\begin{eqnarray}}
\def \eea {\end{eqnarray}}

\def \Tr  {\bf{Tr}}

\begin{document}

\preprint{ECTP-2013-06}

\title{Constant Trace Anomaly as a Universal Condition for the Chemical Freeze-Out}

\author{A.~Tawfik}
\email{a.tawfik@eng.mti.edu.eg}
\email{atawfik@cern.ch}
\affiliation{Egyptian Center for Theoretical Physics (ECTP), MTI University, 11571 Cairo, Egypt}
\affiliation{World Laboratory for Cosmology And Particle Physics (WLCAPP), Cairo, Egypt}

\date{\today}

\begin{abstract}

Finding out universal conditions describing the freeze-out parameters was a subject of various phenomenological studies. In the present work, we introduce a new condition based on constant trace anomaly (or interaction measure) calculated in the hadron resonance gas (HRG) model. Various extensions to the {\it ideal} HRG which are conjectured to take into consideration different types of interactions have been analysed. When comparing HRG thermodynamics to that of lattice quantum chromodynamics, we conclude that the hard-core radii are practically irrelevant, especially when HRG includes all resonances with masses less than $2~$GeV.  It is found that the constant trace anomaly (or interaction measure) agrees well with most of previous conditions.

\end{abstract}

\pacs{74.40.Gh, 05.70.Fh}
\keywords{Trace Anomaly (Interaction measure), Fluctuation phenomena non-equilibrium processes, Phase transitions in statistical mechanics and thermodynamics} 

\maketitle


\section{Introduction}

The QCD trace anomaly which is defined as $(\epsilon-3 p)/T^4$ is also known as the interaction measure. This quantity can be derived from the trace of the energy-momentum tensor, $T_{\mu}^{\mu}=\epsilon-3 p$ and  is conjectured to be sensitive to the presence of massive hadronic states. For instance, for non-interacting hadron gas at vanishing chemical potential
\bea
\frac{\epsilon-3 p}{T^4} &=& \int_0^{\infty} dm \, \rho(m) \int \frac{d^3 p}{(2 \pi)^3}\, \frac{E^2-p^2}{E}\, e^{-\frac{E}{T}} \simeq  \frac{1}{2 \pi^2} \int_0^{\infty} \left(\frac{m}{T}\right)^3 \, \rho(m)  \, K_1\left(\frac{m}{T}\right)\, dm,  \label{eq:Imuel1}
\eea
where $\rho(m)$ is the mass spectrum \cite{hgdrn} which relates the number of hadronic resonances to their masses as an exponential. In general, all thermodynamic quantities are sensitive to $\rho(m)$. For the hadronic resonances which not yet measured in experiments, a parametrization for total spectral weight has been introduced  \cite{brnt}. In the present work, we only include known resonance states with mass $\leq 2~$GeV instead of the Hagedorn mass spectrum \cite{hgdrn}. In the classical limit, the trace anomaly at finite chemical potential $\mu$ reads
\bea
\frac{\epsilon-3 p}{T^4} &\simeq &  \frac{g}{2 \pi^2} \, e^{\mu/T} \, \left(\frac{m}{T}\right)^2 \left[\frac{m}{T}\, K_1\left(\frac{m}{T}\right)\right].
\eea
In addition to these aspects, there are other reasons speak for utilizing even {\it ideal} hadron resonance gas (HRG) model in predicting the hadron abundances and their thermodynamics. As will be discussed, the HRG model seems to provide a good description for the thermal  evolution of the thermodynamic quantities in the hadronic matter~\cite{Tawfik:2004sw,Karsch:2003vd,Karsch:2003zq,Redlich:2004gp,Tawfik:2004vv,Tawfik:2006yq,Tawfik:2010uh,Tawfik:2010pt,Tawfik:2012zz}. Furthermore the HRG model has been successfully utilized in characterizing two different conditions generating the chemical freeze-out at finite densities, namely $s/T^3=7$ \cite{sT3p1,sT3p2,Tawfik:2005qn,Tawfik:2004ss} and $\kappa\, \sigma^2=0$ \cite{Tawfik:2013dba}. As introduced in Ref. \cite{Tawfik:2004ss}, constant $s/T^3$ is accompanied with constant $s/n$.  Recently, the HRG has been used to calculate the higher order  moments of particle multiplicity using a grand canonical partition function of an ideal gas with all experimentally observed states up to a certain large mass as constituents \cite{Tawfik:2012si}. The grand canonical ensemble includes two important features \cite{Tawfik:2004sw}; the kinetic energies and the summation over all degrees of freedom and energies of the resonances. On the other hand, it is known that the formation of resonances can only be achieved through strong interactions~\cite{hgdrn}; {\it Resonances (fireballs) are composed of further resonances (fireballs), which in turn consist of  resonances (fireballs) and so on}. In other words, the contributions of the hadron resonances to the partition function are the same as those of free particles with some effective mass.  At temperatures comparable to the resonance half-width, the effective mass approaches the physical one \cite{Tawfik:2004sw}. Thus, at high temperatures, the strong interactions are conjectured to be taken into consideration through including heavy resonances. It is found that the hadron resonances with masses up to $2\;$GeV include a suitable set of constituents needed for the partition function ~\cite{Karsch:2003vd,Karsch:2003zq,Redlich:2004gp,Tawfik:2004sw,Tawfik:2004vv,Tawfik:2006yq,Tawfik:2010uh,Tawfik:2010pt,Tawfik:2012zz}. Such a way, the singularity expected at the Hagedorn temperature~\cite{Karsch:2003zq,Karsch:2003vd} can be avoided. The strong interactions are assumed to be taken into consideration. In light of this, the validity of HRG is limited to temperatures below the critical one, $T_c$. An extensive discussion on this point will be elaborated in section \ref{sec:extn}.

In high energy experiments, the produced particles and their correlations are conjectured to provide information about the nature, composition and size of the medium from which they are originating. To determine the freeze-out parameters at various center-of-mass energies $\sqrt{s_{NN}}$, the particle yields are analysed in terms of temperature $T$ and baryon chemical potential $\mu$. The baryon density is related to the chemical potential, which is given by the nucleon stopping in the collision region. Furthermore, the chemical freeze-out is defined as the stage in the evolution of the hadronic system when inelastic collisions entirely cease and the relative particle ratios become fixed. Both $T$ and $\mu$ can be related to $\sqrt{s_{NN}}$ \cite{jean2006}. In the present work, we introduce constant trace anomaly (or constant interaction measure) to re-estimate the available sets of $T-\mu$ that were deduced from the different experiments, Fig. \ref{fig:e3p}.

The model is introduced in section \ref{sec:model}. In section \ref{sec:extn}, we discuss different extensions to the ideal hadron gas.  Section \ref{sec:phys} is devoted to introduce the novel condition describing the freeze-out parameters and their dependence on $\sqrt{s_{NN}}$. Physics of constant trace anomaly is presented in section \ref{sec:others}. Other conditions for chemical freeze-out are reviewed in section \ref{sec:rslt}. The conclusions are outlined in section \ref{sec:conc}.

\section{The Hadron Resonance Gas Model}
\label{sec:model}

The hadron resonances treated as a free gas~\cite{Karsch:2003vd,Karsch:2003zq,Redlich:2004gp,Tawfik:2004sw,Tawfik:2004vv} are conjectured to add to the thermodynamic pressure in the hadronic phase (below $T_c$). This statement is valid for free as well as for strongly interacting resonances. 
It has been shown that the thermodynamics of a strongly interacting  system can also be approximated by ideal gas composed of hadron resonances with masses $\le 2~$GeV ~\cite{Tawfik:2004sw,Vunog}. Therefore, in the confined phase of QCD, the hadronic phase, is modelled as a non-interacting gas of resonances. The grand canonical partition function reads
\bea
Z(T, \mu, V) &=& \Tr \left[ \exp^{\frac{\mu\, N-H}{T}}\right],
\eea
where $H$ is the Hamiltonian of the system and $T$ ($\mu$) is the temperature (chemical potential). The Hamiltonian is given by the sum of the kinetic energies of relativistic Fermi and Bose particles. The main motivation of using this Hamiltonian is that it contains all relevant degrees of freedom of confined and  {\it strongly interacting} matter. It includes implicitly the interactions that result in formation of resonances. In addition, it has been shown that this model can offer a quite satisfactory description of the particle production in the heavy-ion collisions. With the above assumptions, dynamics of the partition function can be calculated exactly and apparently expressed as a sum over  {\it single-particle partition} functions $Z_i^1$ of all hadrons and their resonances.
\bea \label{eq:lnz1}
\ln Z(T, \mu_i ,V) &=& \sum_i \ln Z^1_i(T,V) = \sum_i\pm \frac{V g_i}{2\pi^2}\int_0^{\infty} k^2 dk \ln\left\{1 \pm \exp[(\mu_i -\varepsilon_i)/T]\right\},
\eea
where $\varepsilon_i(k)=(k^2+ m_i^2)^{1/2}$ is the $i-$th particle dispersion relation, $g_i$ is
spin-isospin degeneracy factor and $\pm$ stands for bosons and fermions, respectively.

Before the discovery of QCD, a probable phase transition of a massless pion gas to a new phase of matter was speculated \cite{lsm1}. Based on statistical models like Hagedorn \cite{hgdrn1} and statistical Bootstrap model \cite{boots1,boots2}, the thermodynamics of such an ideal pion gas has been studied, extensively. After the QCD, the new phase of matter is now known as the quark gluon plasma (QGP). The physical picture was that at $T_c$ the additional degrees of freedom carried by QGP are to be released resulting in an increase in the thermodynamic quantities like energy and pressure densities. The success of HRG model in reproducing lattice QCD results at various quark flavors and masses (below $T_c$) changed this physical picture drastically. Instead of releasing additional degrees of freedom at $T>T_c$, the hadronic system reduces its effective degrees of freedom, namely at $T<T_c$. In other words, the hadron gas has many degrees of freedom than QGP.

At finite temperature $T$ and baryon chemical potential $\mu_i $, the pressure of the $i$-th hadron or resonance species reads 
\begin{equation}
\label{eq:prss} 
p(T,\mu_i ) = \pm \frac{g_i}{2\pi^2}T \int_{0}^{\infty}\,
           k^2\, dk  \ln\left\{1 \pm \exp[(\mu_i -\varepsilon_i)/T]\right\}.
\end{equation} 
As no phase transition is conjectured in the HRG model, summing over all hadron resonances results in the final thermodynamic pressure (of the hadronic phase). 

The switching between hadron and quark chemistry is given by the relations between  the {\it hadronic} chemical potentials and the quark constituents;  $\mu_i =3\, n_b\, \mu_q + n_s\, \mu_S$, with $n_b$($n_s$) being baryon (strange) quantum number. The chemical potential assigned to the light quarks is given as $\mu_q=(\mu_u+\mu_d)/2$ and the one assigned to strange quark reads $\mu_S=\mu_q-\mu_s$. It is worthwhile to notice that the strangeness chemical potential $\mu_S$ should be calculated as a function of $T$ and $\mu_i $. In doing this, it is assumed that the overall strange quantum number has to remain conserved in the heavy-ion collisions~\cite{Tawfik:2004sw}.

\subsection{Extensions to the ideal hadron gas: hadron interactions}
\label{sec:extn}

\begin{figure}[htb]
\includegraphics[angle=-90,width=8.cm]{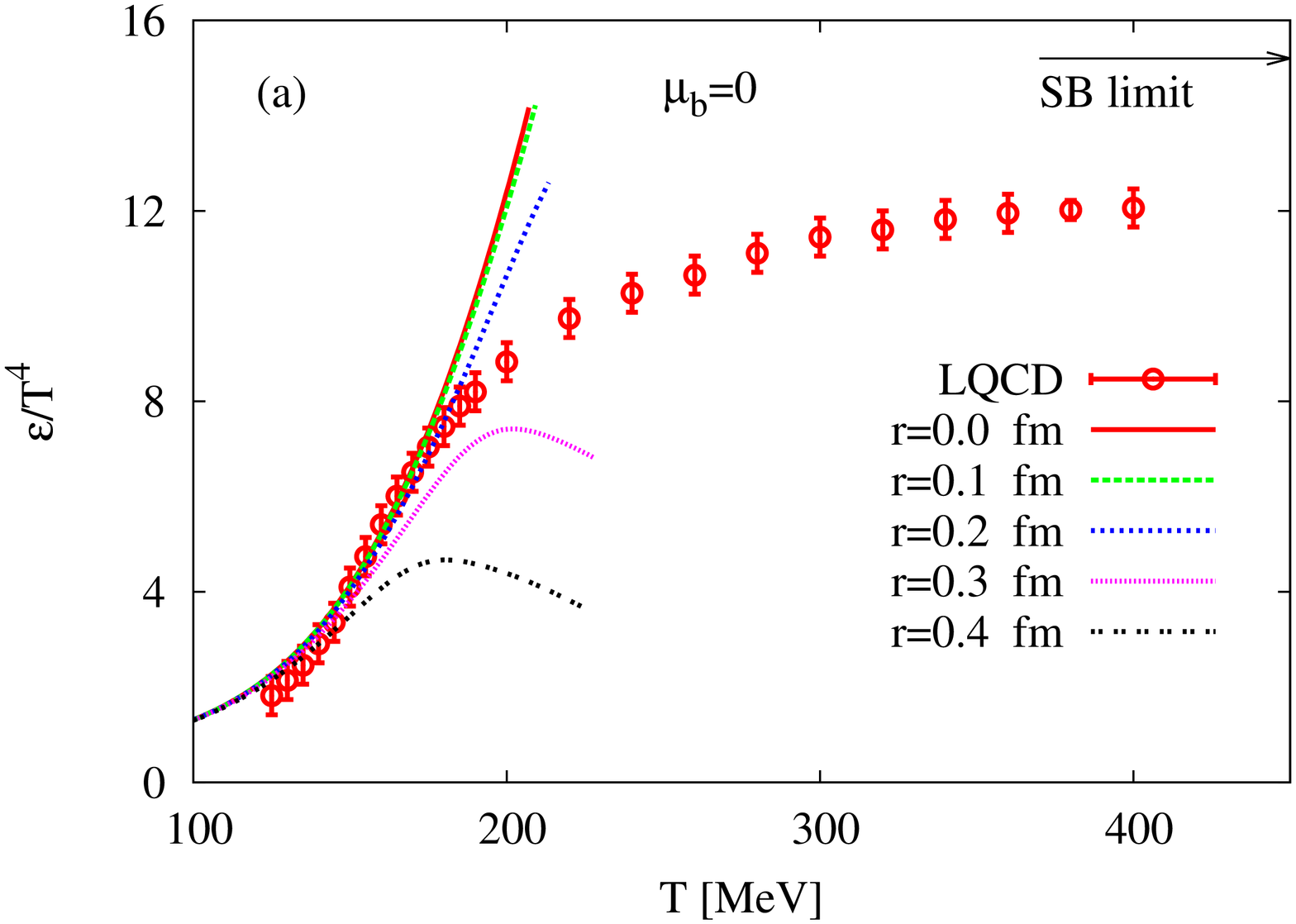}
\includegraphics[angle=-90,width=8.cm]{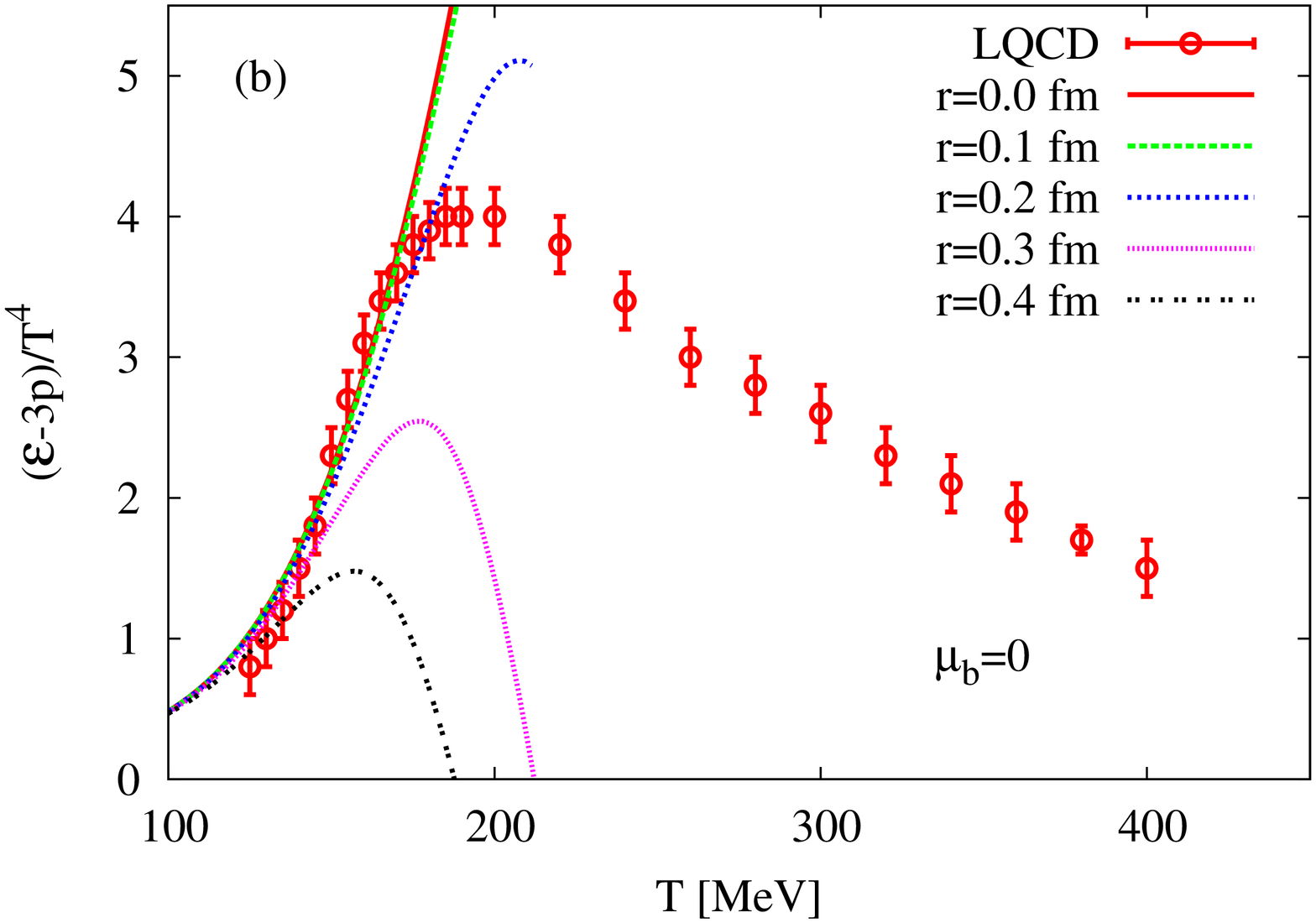}
\caption{\footnotesize (Color online) The normalized energy density is given in dependence on the temperature at vanishing chemical potential in panel (a). The symbols represent the lattice QCD simulations \cite{latFodor}. The curves are the HRG calculations at different hard-core radii of the hadron. Panel (b) presents the same but for the normalized trace anomaly. The results for $T>T_c$ are apparently out of scope of the HRG model. }
\label{fig:vdW} 
\end{figure}

In literature, three types of interactions can be implemented into the {\it ideal} hadron gas. The repulsive interactions between hadrons are considered as a phenomenological extension, which would be exclusively based on van der Waals excluded volume \cite{exclV1,exclV2,exclV3,exclV4}.  Accordingly, considerable modifications in thermodynamics of hadron gas including energy, entropy and number densities are likely. There are intensive theoretical works devoted to estimate the excluded volume and its effects on the particle production and fluctuations \cite{exclV5}, for instance. It is conjectured that the hard-core radius of hadron nuclei can be related to the multiplicity fluctuations \cite{exclV6}. Assuming that hadrons are spheres and  all have the same radius, we compare between different  radii in Fig. \ref{fig:vdW}. On other hand, the assumption that the radii would depend on the hadron masses and sizes could come up with a very small improvement.

The first principle lattice QCD simulations for various thermodynamic quantities offer an essential framework to check the ability of extended {\it ideal} hadron gas, in which the excluded volume is taken into consideration \cite{apj}, to describe the hadronic matter in thermal and dense medium. Figure \ref{fig:vdW} compares normalized energy density and trace anomaly as calculated in lattice QCD and HRG model. The symbols with error bars represent the lattice QCD simulations for $2+1$ quark flavors with physical quark masses in continuum limit, i.e. vanishing lattice spacing \cite{latFodor}. The curves are the HRG calculations at different hard-core radii of hadron resonances, $r$. We note that increasing the hard-core radius reduces the ability to reproduce the lattice QCD results. Two remarks are now in order. At $0\leq r<0.2~$fm, the ability of HRG model to reproduce the lattice energy density or trace anomaly is apparently very high. Furthermore, we note that varying $r$ in this region makes almost no effect i.e., the three radii, $r=[0.0,0.1,0.2]~$fm, have almost the same results. At $r>0.2~$fm, the disagreement  becomes obvious and increases with increasing $r$. At higher temperatures, the resulting thermodynamic quantities, for instance energy density and trace anomaly become {\it non}-physical. For example, the energy density and trance anomaly nearly tends toward vanishing. So far, we conclude that the excluded volume is practically irrelevant. It comes up with a negligible effect, at $r\leq 0.2~$fm. On other hand, a remarkable deviation from the lattice QCD calculations appears, especially when relative large values are assigned to $r$. With this regard, it has to be taken into consideration that the excluded volume itself is conjectured to assure the thermodynamic consistency in the HRG model \cite{exclV1,exclV2,exclV3,exclV4,exclV5,exclV6}. 

It is obvious that the thermodynamic quantities calculated from the HRG model are likely to diverge at $T_c$ \cite{hgdrn,hgdrn1}. It is a remarkable finding that despite the mass cut-off at $2~$GeV,  the energy density remains finite even when $T$ exceeds $T_c$. Apparently, this is the main reason why the trace anomaly gets negative values. The correction to pressure is tiny or negligible  \cite{exclV1,exclV2,exclV3,exclV4,exclV5,exclV6}.  Nevertheless, the finite hard-core should not be believed to reproduce the lattice QCD simulations at $T>T_c$. The validity of HRG model is strictly limited to $T<T_c$.

The second type of extensions has been introduced by Uhlenbeck and Gropper \cite{uhlnb}. This is mainly the correlation. The {\it non}-ideal (correlated) hadron statistics is given by the classical integral if the Boltzmann factor $\exp(-\phi_{ij}/T)$ is corrected as follows. 
\bea \label{eq:uhlb}
\exp\left(-\frac{\phi_{ij}}{T}\right)\left[1\pm e^{-m\, T\, r_{i j}^2}\right],
\eea
where $r_{ij}$ is the average correlation distance between $i$-th and $j$-th particle. $\phi_{i j}$ is the interaction potential between $i$-th and $j$-th particle pairs. Apparently, the summation over all pairs gives the total potential energy. This kind of modifications takes into account correlations and also the {\it non-ideality} of the hadron gas. The latter would among others refer to the discreteness of the energy levels. Uhlenbeck and Gropper  introduced an additional correction but concluded that it is only valid at very low temperatures \cite{uhlnb}. It should be noticed that the correction, expression (\ref{eq:uhlb}), is belonging to {\it generic} types of correlation interactions.

As introduced in Ref. \cite{Tawfik:2004sw}, the third type of interactions to be implemented in the ideal hadron gas is the attraction. The Hagedorn states are considered as the framework to study the physics of strongly interacting matter for temperatures $<T_c$. The hadron resonances add attraction interaction to the partition function. In other words, the Hagedorn interaction finds its description in the hadron mass spectrum, $\rho(m)$, Eq. (\ref{eq:Imuel1}). Using the hadrons and resonances which are verified experimentally, the limits of the exponential limit of $\rho(m)$ can be determined. For instance, the mass cut-off may vary from strange to non-strange states Ref. \cite{brnt}, for strange $1.5~$GeV and for non-strange $2.0~$GeV or from bosons to fermions, etc. According to the statistical Bootstrap model \cite{boots1,boots2}, the fireballs are treated as hadronic massive states possessing all conventional hadronic properties. It is apparent that the fireball mass is determined by the mass spectrum. Furthermore, fireballs are consisting of further fireballs. This is only valid at the statistical equilibrium of an ensemble consisting of an undetermined number of states (fireballs).

On one hand, the whole spectrum of possible interactions can be represented by $S$-matrix.  According to \cite{Tawfik:2004sw}, the fugacity term can be expanded to include various kinds of interactions. In such a way, the $S$-matrix would give the plausible scattering processes taking place in the system of interest. This would be partly understood that including hadron resonances with some effective masses has almost the same effect as that of a free particle with same mass. At high energy, the effective mass approaches the physical value. In other words, even strong interactions are taken into consideration via heavy resonances. These conclusions suggest that the grand canonical partition function is able to simulate various types of interactions, when hadron resonances with masses up to $2~$GeV are included. As discussed, this sets the limits of Hagerdorn mass spectrum, Eq. (\ref{eq:Imuel1}). The mass cut-off  avoids the Hagedorn's singularity.  A conclusive convincing proof has been presented through confronting HRG to lattice QCD results  \cite{Karsch:2003vd,Karsch:2003zq,Redlich:2004gp,Tawfik:2004sw,Tawfik:2004vv}. Figure \ref{fig:vdW} illustrates the excellent agreement between HRG with $r=0~$fm and the lattice QCD calculations. It should be noticed that the results at $T>T_c$ are out of scope of the HRG model. 

So far, we conclude that the attraction interaction is very sufficient to overcome the hard-core repulsion interaction. In light of this, we comment on the conclusion of Ref. \cite{apj}. In framework of  interacting hadron resonance gas, a thermal evaluation of thermodynamic quantities has been proposed. The interactions to be implemented into HRG model are mainly van der Waals repulsion which are implemented through correction for the finite size of hadrons. Different values for the hadron radii can be assigned to the baryons and mesons. The authors studied the sensitivity of the modified HRG model calculations of  hadron radii. The results on different thermodynamic quantities were confronted with predictions from lattice QCD simulations. Therefore, the conclusion would be understood, as  hadron resonances with masses up to $3~$GeV are taken into consideration. At this mass cut-off, the exponential description of the mass spectrum would be no longer valid. Furthermore, it is straightforward to deduce from textbook that heavier masses lead to lower thermodynamic quantities. It is correctly emphasized \cite{lFodor10} that including all known hadrons up to $2.5$ or even $3.0$ GeV would increase the number of hadron resonances by a few states with masses $>2~$GeV. An attempt to improve the HRG model by including an exponential mass spectrum for these very heavy resonances has been proposed \cite{brnt}. In Refs. \cite{lFodor10,apj} only known states and not the mass spectrum are taken into account. For this reason, the authors of \cite{apj} concluded that the HRG model with small or even vanishing radii gives thermodynamic quantities which apparently are less steeply than in case of a {\it ideal} HRG.

It is apparent that the excluded volume has almost no effect, especially when the baryon chemical potential is small. Also, the first principle lattice QCD calculations are reliable at small baryon chemical potential. On the other hand, the statistical-thermal models would make no sense, if the hadron resonances are point-like and the baryon chemical potential is large.

\section{Results}
\label{sec:rslt}

The HRG calculations are performed as follows. Starting with a certain value of the baryon chemical potential $\mu_b$, the temperature $T$ is increased very slowly. At this value of $\mu_b$ and at each raise in $T$, the strangeness chemical potential $\mu_S$ is determined under the condition that the strange quantum numbers should remain conserved in the heavy-ion collisions. Having the three values of $\mu_b$, $T$  and $\mu_S$, then all thermodynamic quantities  including  $(\epsilon-3p)/T^4$ are calculated. When the trace anomaly reaches the value $7/2$, then the temperature $T$ and chemical potential $\mu_b$ are registered. This procedure is repeated over all values of $\mu_b$. It is worthwhile to mention that the normalized trace anomaly is calculated using grand canonical statistics.   Furthermore, in HRG calculations no statistical fitting has been applied in determining any thermodynamic quantities, including pressure, entropy density and trace anomaly.  

\begin{figure}[htb]
\includegraphics[angle=-90,width=14cm]{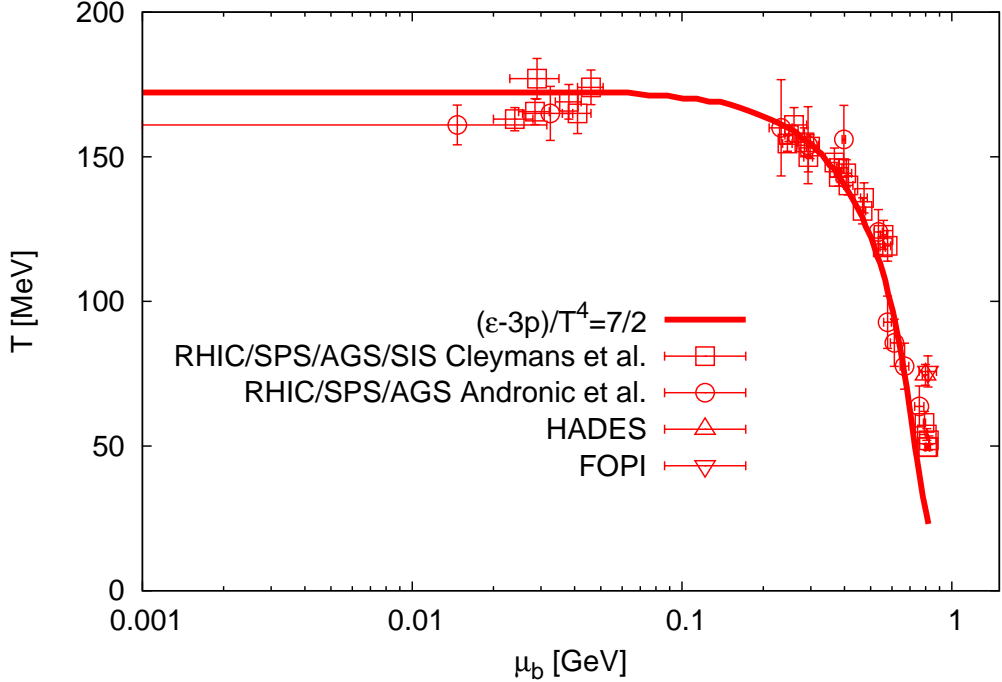}
\caption{\footnotesize (Color online) The freeze-out parameters ($T$ vs. $\mu_b$) are compared with constant trace anomaly calculated in HRG model (solid curve). Symbols with error bars are the phenomenologically estimated results (see text). }
\label{fig:e3p} 
\end{figure}

In Fig. \ref{fig:e3p}, the freeze-out parameters, temperature $T$ and baryon chemical potential $\mu_b$, as calculated in the HRG model are plotted in a log-log graph (solid curve). The symbols with error bars represent the phenomenologically estimated parameters know as experimental data \cite{jean2006,dataCR}. The experimental data covers a center-of-mass energy ranging from couple GeV in SchwerIonen Synchrotron (SIS) to several TeV, in Large Hadron Collider (LHC). HADES \cite{hades} and FOPI \cite{fopi} results are also included.  

The phenomenological parameters \cite{jean2006,dataCR} have been estimated as follows. In different high-energy experiments (corresponding to different center-of-mass energies), various particle ratios measured in these experiments are re-calculated by the thermal models \cite{jean2006,dataCR}. In doing this, the chemical potential $\mu_b$ measured from the stopping power is used as an input. In light of this, the thermal models are used to merely estimate the freeze-out temperature. Figure \ref{fig:e3p} presents the freeze-out diagram relating $T$ to $\mu_b$ (symbols with error bars). In the present paper, a new universal description is suggested. It is assumed that constant trace anomaly is able to reproduce the freeze-out diagram, $T$ vs. $\mu_b$.

\section{Physics of constant trace anomaly or constant interaction measure}
\label{sec:phys}

The QCD equation of state can be deduced from the energy-momentum tensor, for instance, the normalized pressure can be obtained for the integral of $(\epsilon-3 p)/T^5$.  For completeness, we mention that the trace anomaly gets related the QCD coupling constant so that $I(T)/T^4\propto T^4\, \alpha_s^2$ \cite{peter}, where $I(T)=\epsilon(T)-3 p(T)$. In light of this, essential information about weakly coupled systems would be provided through trace anomaly. 

A universal parametrization for the QCD trace anomaly at $\mu=0$ was proposed \cite{FodorFit1}
\bea \label{eq:parmQCD}
\frac{I(T,\mu_b=0)}{T^4} &=& e^{-\frac{h_1}{t}-\frac{h_2}{t^2}} \left\{h_0+\frac{f_0 \left[\tanh(f_1 t+f_2)\right]+1}{1+g_1 t+g_2 t^2}\right\},
\eea
where $t=T/200$. The fitting parameters are listed in Ref. \cite{FodorFit1}. This gives the thermal evolution of $I(T)$. A parametrization at finite  $\mu$ was suggested \cite{OweRev}
\bea
\label{eq:OweP1}
\frac{I(T,\mu_b)}{T^4} &=& \frac{I(T,\mu_b=0)}{T^4} + \frac{1}{2} \frac{\mu_b^2}{T} \frac{\partial \chi_2(T,\mu_b)}{\partial T},
\eea
where the susceptibility $\chi_2(T,\mu_b)$  is given by the second derivative of the partition function, 
\bea
\chi_2(T,\mu_b) &=& \frac{1}{T}  \frac{\partial^2 \ln Z(T,\mu_b)}{\partial^2 \mu_b}.
\eea
In the classical limit, the derivative of $\chi_2(T,\mu_b)$ wrt temperature reads
\bea
\label{eq:My1}
\left.\frac{\partial \chi_2(T,\mu_b)}{\partial T}\right|_{\mu_b=0} &=& \frac{\chi_2(T,\mu_b)}{T^2} - \frac{g}{4\pi^2}\; e^{\mu_b/T}\; T \left(\frac{m}{T}\right)^3  \left[K_1\left(\frac{m}{T}\right) + K_3\left(\frac{m}{T}\right)\right].
\eea
When implementing the result in Eq. (\ref{eq:OweP1}), we get 
\bea \label{eq:Icls}
\frac{I(T,\mu_b)}{T^4} &=& \frac{I(T)}{T^4} - \frac{\chi_2(T,\mu_b)}{T^2}\, \frac{\mu_b^3}{2 T} + \frac{g}{8 \pi^2}\; e^{\mu_b/T}\; \mu_b^2 \left(\frac{m}{T}\right)^3  \left[K_1\left(\frac{m}{T}\right) + K_3\left(\frac{m}{T}\right)\right].
\eea

When assuming that $I(T,\mu_b)/T^4=7/2$ at vanishing and finite $\mu_b$, then
\bea
\frac{1}{2}  &=&  T \left(\frac{m}{T}\right) \, K_2\left(\frac{m}{T}\right) \, \left[K_1\left(\frac{m}{T}\right) + K_3\left(\frac{m}{T}\right)\right]
\eea

In grand canonical ensemble at  $T<T_c$
\bea
\label{eq:OweP3}
\frac{I(T,\mu_b)}{T^4}  &=& \frac{I(T,\mu_b=0)}{T^4} \pm \frac{1}{2}\left(\frac{\mu_b}{T}\right)^2\; \frac{\chi_{2}(T,\mu_b)}{T^2} + \nonumber \\
 &&
 \frac{1}{2}\left(\frac{\mu_b}{T}\right)^2 \frac{g}{2 \pi ^2} \frac{1}{T^2}\int _0^{\infty }  \left( 
 \frac{2 e^{-\frac{3 \sqrt{m^2+p^2}}{T}+\frac{3 \mu_b}{T}} \left(\frac{\sqrt{m^2+p^2}}{T^2}-\frac{\mu_b}{T^2}\right)}{\left(1\pm e^{-\frac{\sqrt{m^2+p^2}}{T}+\frac{\mu_b}{T}}\right)^3} - \right. \nonumber\\
 &&
 3\frac{e^{-\frac{2 \sqrt{m^2+p^2}}{T}+\frac{2 \mu_b}{T}} \left(\frac{\sqrt{m^2+p^2}}{T^2}-\frac{\mu_b}{T^2}\right) }{\left(1\pm e^{-\frac{\sqrt{m^2+p^2}}{T}+\frac{\mu_b}{T}}\right)^2 } +
 \left.\frac{e^{-\frac{\sqrt{m^2+p^2}}{T}+\frac{\mu_b}{T}} \left(\frac{\sqrt{m^2+p^2}}{T^2}-\frac{\mu_b}{T^2}\right)}{\left(1\pm e^{-\frac{\sqrt{m^2+p^2}}{T}+\frac{\mu_b}{T}}\right) }\right) p^2 dp, \nonumber \\
&=& \frac{I(T,\mu_b=0)}{T^4} \pm \frac{1}{2}\left(\frac{\mu_b}{T}\right)^2\; \frac{\chi_{2}(T,\mu_b)}{T^2} + \nonumber \\
&& \frac{1}{2}\left(\frac{\mu_b}{T}\right)^2 \frac{g}{2 \pi ^2} \frac{1}{T^4}\, \int _0^{\infty}  e^{\mu_b/T} \left[\varepsilon-\mu_b\right] \frac{F(T,\mu_b)}{\left(e^{\mu_b/T} \pm e^{\varepsilon/T}\right)^3}  p^2 dp,
\eea
where
\bea
F(T,\mu_b) &=& \left\{ 
 \begin{array}{l l} 
 2 e^{2 \mu_b/T} + e^{(\varepsilon+\mu_b)/T} + e^{2 \varepsilon/T} & \text{for bosons}\\
  & \\
  3 e^{(\varepsilon+\mu_b)/T} - 4 e^{2 \mu_b/T} - e^{2 \varepsilon/T} & \text{for fermions} \\
 \end{array} 
\right..  \nonumber
\eea
The coefficient of $(\mu_b/T)^2$ seems to play a crucial role in estimating the chemical parameters, $T$ and $\mu_b$, the freeze-out diagram. The second term of Eq. (\ref{eq:OweP3}) can be decomposed into bosonic and fermionic parts
\bea
\pm\frac{1}{2}\left(\frac{\mu_b}{T}\right)^2\; \frac{\chi_{2}(T,\mu_b)}{T^2} &=& \frac{1}{2}\left(\frac{\mu_b}{T}\right)^2\; \left[ \frac{\chi^{(B)}_{2}(T)}{T^2} - \frac{\chi^{(F)}_{2}(T,\mu_b)}{T^2}\right],
\eea
revealing that the fermionic susceptibility  is to a large extend responsible for the $T-\mu_b$ curvature.

\section{Other conditions for the chemical freeze-out}
\label{sec:others}

Starting from phenomenological observations at SIS energy, it was found that the averaged energy per averaged particle $\epsilon /  n \approx 1~$GeV \cite{jeanRedlich}, where Boltzmann approximations are applied, this constant ratio is assumed to describe the whole $T-\mu_b$ diagram. For completeness, we mention that the authors assumed that the pions and rho-mesons are conjectured to get dominant at high $T$ and small $\mu_b$. The second criterion assumes that total baryon number density $ n_b +  n_{\bar{b}} \approx 0.12~$fm$^{-3}$ \cite{nb01}. In framework of percolation theory, a third criterion has been suggested \cite{percl}. As shown in Fig. 2 of \cite{Tawfik:2005qn}, the last two criteria seem to give almost identical results. Both of them are apparently stemming from phenomenological observations. A fourth criterion based on lattice QCD simulations was introduced in Ref.  \cite{Tawfik:2005qn,Tawfik:2004ss}. Accordingly, the entropy normalized to cubic temperature is assumed to remain constant over the whole range of baryon chemical potentials, which is related to the nucleus-nucleus center-of-mass energies $\sqrt{s_{NN}}$ \cite{jean2006}. An extensive comparison between constant $\epsilon /  n$ and constant $s/T^3$ is given in \cite{Tawfik:2005qn,Tawfik:2004ss}. 

In the HRG model, the thermodynamic quantities generating the chemical freeze-out are deduced \cite{Tawfik:2005qn,Tawfik:2004ss}. The motivation of suggesting constant normalized entropy is the comparison to the lattice QCD simulations with two and three quark flavors. We found the $s/T^3=5$ for two flavors and $s/T^3=7$ for three flavors. Furthermore, we confront the hadron resonance gas results to the experimental estimation for the freeze-out parameters, $T$ and $\mu_b$. 

Another novel condition characterizing the freeze-out parameters has been introduced in Ref. \cite{Tawfik:2013dba}. To this extend, the higher order moments are applied \cite{Tawfik:2012si}. Vanishing ${\kappa}\, \sigma^2$ or equivalently $m_4/\chi=3$ results in $T$ - $\mu_b$ sets coincident with the phenomenologically estimated ones.  Recently, lattice QCD calculations confirm the same connection between the ratios of higher order fluctuations and the freeze-out parameters \cite{fodorFO,nakamura1}. 

\begin{figure}[htb]
\includegraphics[angle=-90,width=14cm]{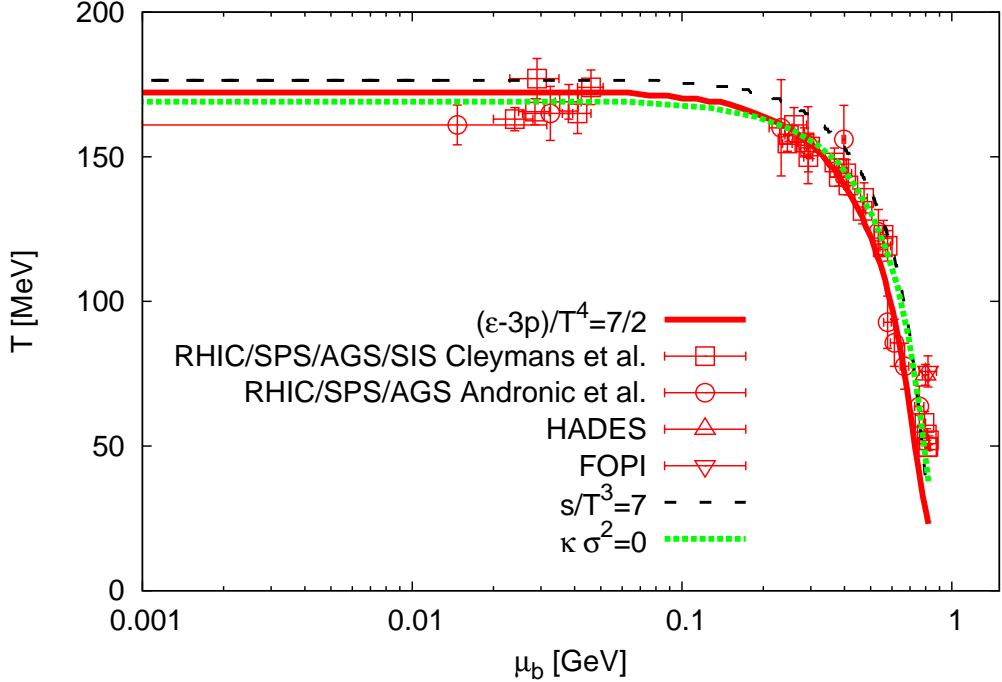}
\caption{\footnotesize (Color online) The same as in Fig. \ref{fig:e3p} with an additional comparison with  $s/T^3=7$  \cite{Tawfik:2005qn,Tawfik:2004ss} and vanishing ${\kappa}\, \sigma^2$ \cite{Tawfik:2013dba}. The agreement is very convincing. }
\label{fig:comp} 
\end{figure}

Figure \ref{fig:comp} compares between the present condition, $(\epsilon-3 p)/T^4=7/2$ and two other conditions, namely $s/T^3=7$  \cite{Tawfik:2005qn,Tawfik:2004ss} and ${\kappa}\, \sigma^2=0$ \cite{Tawfik:2013dba}. The agreement is convincing.

\section{Conclusions}
\label{sec:conc}

So far, we conclude that the freeze-out parameters deduced at various center-of-mass energies are well reproducible  in the HRG model using the condition that the trace anomaly (interaction measure) $(\epsilon-3 p)/T^4$ remains constant. We found that $(\epsilon-3 p)/T^4=7/2$ reproduces very well the freeze-out diagram. From Eq. (\ref{eq:Icls}), constant normalized $I(T,\mu_b)=I(T)$ leads to
\bea
\chi_2(T,\mu_b)  &=& \frac{g}{4 \pi^2}\; e^{\mu_b/T}\;  \frac{m^3}{\mu_b}\; \left[K_1\left(\frac{m}{T}\right) + K_3\left(\frac{m}{T}\right)\right]. 
\eea
Then, the chemical freeze-out potential 
\bea
\mu_b &=& m\; \frac{K_1\left(\frac{m}{T}\right) + K_3\left(\frac{m}{T}\right)}{2 K_2\left(\frac{m}{T}\right)}.
\eea
seems to be related to the effective mass, $m$. 

Furthermore, we conclude the present condition, $(\epsilon-3 p)/T^4=7/2$ agrees with two other conditions, namely $s/T^3=7$  \cite{Tawfik:2005qn,Tawfik:2004ss} and ${\kappa}\, \sigma^2=0$ \cite{Tawfik:2013dba}.

\section*{Acknowledgements}

The author likes to thank Prof. Antonino Zichichi for his kind invitation to attend the International School of Subnuclear Physics 2013 at the ”Ettore Majorana Foundation and Centre for Scientific
Culture” in Erice-Italy, where the present script was completed.



\end{document}